\documentclass[10pt, paper=a4, UKenglish]{article}
\usepackage{graphicx}
\usepackage{amsmath}
\usepackage{upgreek}
\def\Title#1{\begin{center} {\Large #1 } \end{center}}
\def\Author#1{\begin{center}{ \sc #1} \end{center}}
\def\Address#1{\begin{center}{ \it #1} \end{center}}

\newcommand\pubblock{\rightline{\begin{tabular}{l} Proceedings of the CTD 2025\\ \pubnumber\\
         \pubdate  \end{tabular}}}

\newenvironment{Abstract}{\begin{quotation} \begin{center} 
             \large ABSTRACT \end{center}\bigskip 
      \begin{center}\begin{large}}{\end{large}\end{center} \end{quotation}}

\newenvironment{Presented}{\begin{quotation} \begin{center} 
             PRESENTED AT\end{center}\bigskip 
      \begin{center}\begin{large}}{\end{large}\end{center} \end{quotation}}

\def\beq{\begin{equation}}
\def\eeq#1{\label{#1}\end{equation}}
\def\eeqn{\end{equation}}

\def\beqa{\begin{eqnarray}}
\def\eeqa#1{\label{#1}\end{eqnarray}}
\def\eeqan{\end{eqnarray}}

\let\bar=\overbar

\def\Dslash{\not{\hbox{\kern-4pt $D$}}}
\def\dslash{\not{\hbox{\kern-2pt $\del$}}}

\def\msb{{\bar{\ssstyle M \kern -1pt S}}}

\usepackage{color}
\usepackage{lineno}
\usepackage{subfig}
\usepackage{hyperref}

\newcommand\pubnumber{PROC-CTD2025-010}

\newcommand\pubdate{\today}

\def\affiliation{
CERN \\
Esplanade des Particules 1\\
1217 Meyrin, Switzerland}

\newcommand{\conference}{Connecting the Dots Workshop (CTD 2025)\\
November 10-14, 2025}

\usepackage{fancyhdr}
\definecolor{mygrey}{RGB}{105,105,105}
\begin{document}


\large
\begin{titlepage}
\pubblock

\vfill
\Title{Improvements of the ALICE GPU TPC tracking and GPU framework for online and offline processing of Run 3 Pb--Pb data}
\vfill

\Author{David Rohr on behalf of the ALICE Collaboration}
\Address{\affiliation}
\vfill

\begin{Abstract}
ALICE is the dedicated heavy ion experiment at the LHC at CERN and records lead-lead collisions at a rate of up to 50 kHz in LHC Run 3.
To cope with such collision and data rates, ALICE uses a new GEM TPC with continuous readout and a GPU-based online computing farm for data compression.
Operating the first GEM TPC of this size with large space charge distortions due to the high collision rate has many implications for the track reconstruction algorithm, both anticipated and unanticipated.
With real Pb--Pb data available, the TPC tracking algorithm needed to be refined, particularly with respect to improved cluster attachment at the inner TPC region.
In order to use the online computing farm efficiently for offline processing when there is no beam in the LHC, ALICE is currently running TPC tracking on GPUs also in offline processing.
For the future, ALICE aims to run more computing steps on the GPU, and to use other GPU-enabled resources besides its online computing farm.
These aspects, along with better possibilities for performance optimizations led to several improvements of the GPU framework and GPU tracking code, particularly using Run Time Compilation (RTC).
The talk will give an overview of the improvements for the ALICE tracking code, mostly based on experience from reconstructing real Pb--Pb data with high TPC occupancy.
In addition, an overview of the online and offline processing status on GPUs will be given, and an overview of how RTC improves the ALICE tracking code and GPU support.
\end{Abstract}

\vfill

\begin{Presented}
\conference
\end{Presented}
\vfill
\end{titlepage}
\def\thefootnote{\fnsymbol{footnote}}
\setcounter{footnote}{0}
%

\normalsize 


\section{Introduction}
\label{intro}

ALICE (A Large Ion Collider Experiment)~\cite{bib:alice} is an experiment dedicated to the study of heavy ion collisions at the LHC (Large Hadron Collider) at CERN.
During the LHC Long Shutdown 2 ALICE upgraded its main tracking detectors~\cite{bib:ls2upgrade} and its computing scheme~\cite{bib:o2tdr} increasing the recording rate of Pb--Pb collisions to 50 kHz.
This is 2 orders of magnitude more data than in LHC Runs 1 and 2.
In particular, the ITS (Inner Tracking System with 7 layers of silicon pixels) and the TPC (Time Projection Chamber) were upgraded.
Rather than using a trigger, all data are read out and stored after being compressed online in a GPU-enabled online computing farm called EPN (Event Processing Nodes).
This is made possible by switching the TPC technology to GEM-based (Gas Electron Multiplier) amplification enabling continuous readout.
When there is no beam in the LHC, the EPN farm is used together with the Grid\footnote{https://home.cern/science/computing/grid - The Worldwide LHC Computing Grid (WLCG)} for the offline reprocessing of the data.
Through the common software framework for online and offline processing, ALICE leverages the EPN GPUs to speed up offline event reconstruction~\cite{bib:chep2023}.

The detector with the largest data volume is the TPC, making the compression of TPC data the most important task of online computing.
The TPC compression is based on the rejection of clusters not used for physics analysis, and on track model compression to reduce the entropy~\cite{bib:ctd2019}, requiring full TPC tracking.
In contrast, online calibration only requires tracks for $\sim$1\% of the data.
Thus, TPC tracking is the most computing-intensive task of online processing, and the full TPC processing from raw decoding over clusterization and tracking until the compression runs fully on GPUs.
In average 5 Pb--Pb collisions and many more pp collisions will occur during a TPC drift time of $\sim$100\;$\upmu$s.
These collisions are overlapping and due to the continuous, triggerless readout, hits cannot be assigned to collisions before tracking.
Collisions or bunch crossings cannot be processed individually, but the fundamental data unit for reconstruction is a time frame of currently 32 LHC orbits equaling~$\sim$2.8\;ms.
The final online compression outputs are CTFs (Compressed Time Frames) with~$\sim$2.8\;ms of data.

\section{ALICE Run 3 TPC Tracking Performance}
\label{section1}

In order to understand the analysis of the tracking performance and the improvements to the tracking algorithm described further below, a quick summary of the ALICE TPC tracking is given.
The TPC is split into two halves, and both halves are further subdivided into 18 trapezoidal sectors each.
The TPC tracking of Run 3 was derived from the ALICE HLT (High Level Trigger)~\cite{bib:hlt} TPC tracking algorithm of Run 2.
It consists of three phases: sector tracking~\cite{bib:tns}, track merging and track fit.

The sector tracking starts with a combinatorial seeding based on a cellular automaton, which will usually find multiple seeds for the same track.
Next, the track following fits all seeds and propagates them through the TPC volume using the Kalman filter picking up more clusters of the track~\cite{bib:tns}.
Tracks are found multiple times, usually with slightly different cluster attachment.
Finally, the ambiguity solving step identifies the best instance of each track, assigns all clusters to this instance and removes the clones.
Due to the many seeds, all steps until here must be fast and cannot be overly complex.
A configurable fraction of a track's clusters, by default 10\%, are allowed to be shared with other tracks to avoid gaps in crossing tracks in high-density regions particularly in the inner TPC pad rows.

The track merging merges tracks first inside the TPC sectors, then between adjacent sectors of the TPC halves, and then across the central electrode.
By design, the sector tracking cannot find very short track segments.
Therefore, tracks ending close to the upper or lower end of the TPC at a sector edge are propagated to the adjacent sector before the merging and extra track following is performed.
Finally, an afterburner merges the individual legs\footnote{Parts of the helix trajectory that are alternatingly inwards and outwards in radial direction.}of low~$p_{\text{T}}$ looping tracks based on the helix parameters, if the consecutive legs have not already been merged by the in-sector merging.

The track fit is the most computationally intensive phase of the TPC tracking and calculates the final track parameters.
It performs three iterations, first inward, then outward, then inward for two reasons:
For stability reasons, the initial parameters are used for the linearization in the Kalman filter.
This makes the fit rely on good initial parameters and a repetition improves the fit result.
In addition, the three-iteration fit allows for a more elaborate cluster-rejection, which is explained in the following.
Rejecting clusters based on a~$\chi^2$ cut during one iteration cannot take the full track information into account and is particularly unstable at the beginning of the iteration when the uncertainties are still large.
Therefore ALICE uses an interpolation approach.
The first, inward iteration guarantees that the following iterations will operate with good initial parameters.
The second, outward iteration stores the current track position and covariance at every TPC pad row in reduced precision to a local variable.
In the final, inward iteration, at each TPC pad row the cluster~$\chi^2$ cut is performed against the interpolation between the current position and covariance with the values stored during the second iteration for this pad row.
Performing interpolation and rejection in the same iteration has the side effect that the removal of one cluster changes the interpolated values of lower pad rows but not of higher pad rows.
Unfortunately, this can lead to a ``domino effect'', where in case of bad parameters mostly in the early phase of the fit, the rejection of one cluster prevents pulling the Kalman state towards the real trajectory increasing the~$\chi^2$ for all following clusters.
Repeating the full interpolation after each cluster rejection would be preferable but repeating the full fit every time would increase the compute time by a large factor.
Particularly for the HLT, where this algorithm originates, this was infeasible and unnecessary.
It should be noted that by design, only cluster rejection is possible but not late attachment after the initial track following.

The TPC GPU tracking has been running successfully since the start of the ALICE Run 3 data taking in 2022 and has met all demands in terms of processing speed online and offline~\cite{bib:chep2023}.
In contrast to HLT times, it has to meet full offline processing physics performance requirements.
Unfortunately at the time of writing, the TPC reconstruction performance is still impacted by non-final correction maps for TPC Space Charge Distortions, but it has become clear that some performance issues come from TPC tracking itself.

\begin{figure}[!htb]
  \centering
  \subfloat[]{\includegraphics[width=0.475\linewidth]{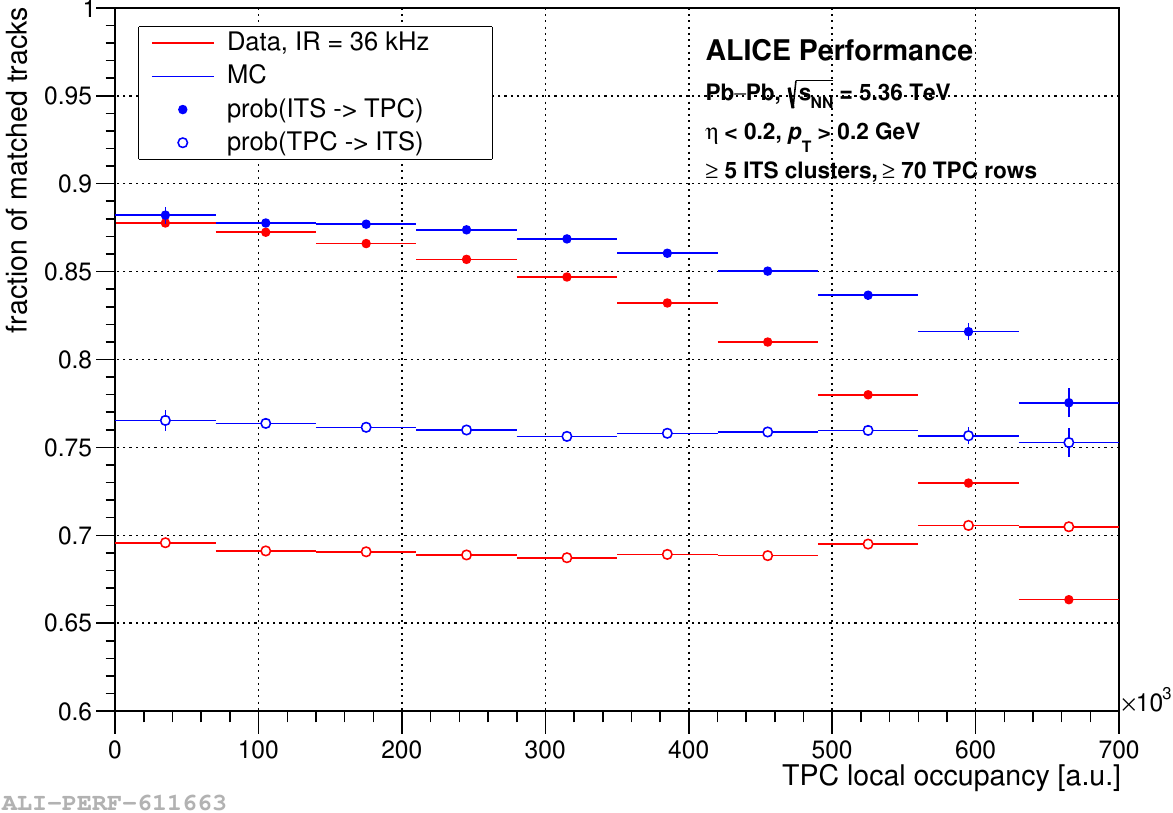}}
  \qquad
  \subfloat[]{\includegraphics[width=0.475\linewidth]{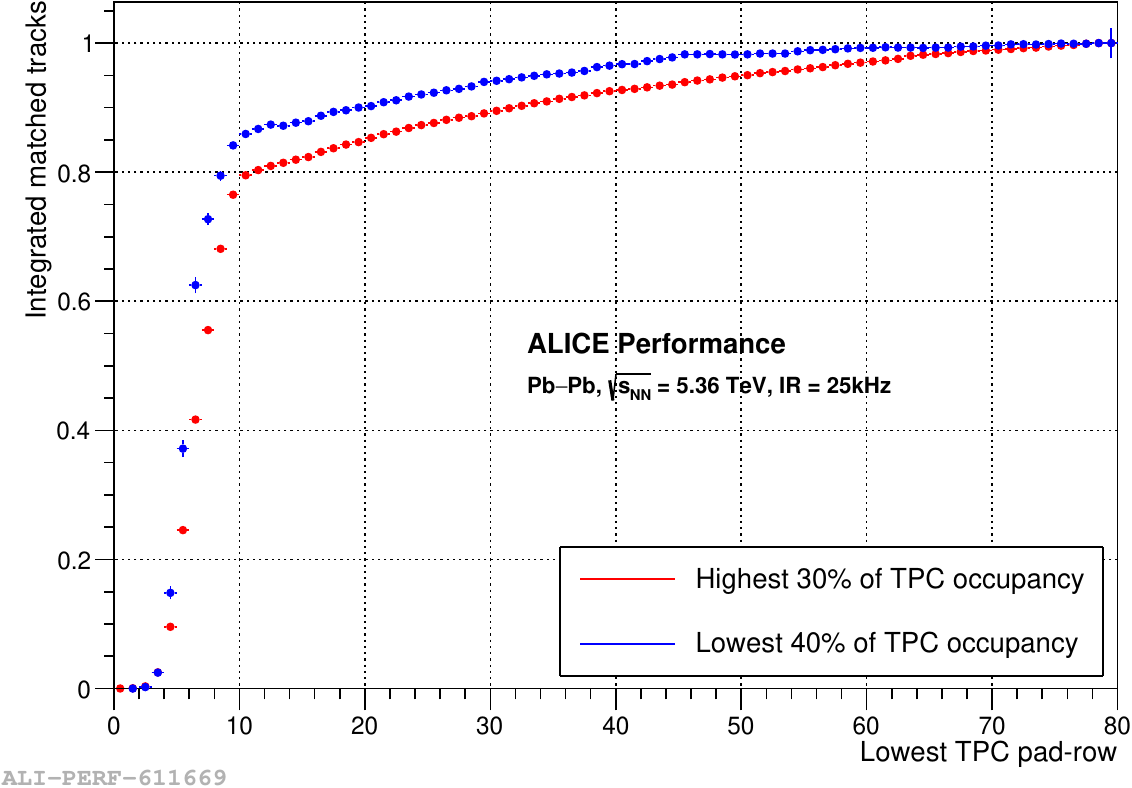}}
  \caption{
    Matching Efficiency for ITS and TPC tracks versus local TPC occupancy.
    (a) shows two curves for processing of raw data and for MC data.
    Solid markers indicate the probability to find a TPC track for a given ITS track.
    Hollow markers indicate in the other way around the probability to find an ITS track for a given TPC track.
    Local TPC occupancy is the number of clusters in the TPC time bins close to the track.
    A cut of~$\lvert \eta \rvert < 0.2$ is applied to show only tracks where all clusters of the track see more or less the same local occupancy.
    (b) shows the probability to find a TPC track for a given ITS track versus the lowest TPC pad row, which has a cluster assigned to the TPC track.
    Two curves are shown for collisions in the highest 30\% and in the lowest 40\% of TPC occupancy.}
  \label{fig:tpcits}
\end{figure}

Fig.~\ref{fig:tpcits} (a) shows the matching efficiency of ITS tracks to TPC tracks.
With higher local TPC occupancy, the chance to find a matching TPC track for a given ITS track decreases.
Currently, this effect is much stronger than anticipated degrading the physics performance.
The effect is visible with data and with MC, while it is not yet understood why the effect is more pronounced in data.
The fact that the matching efficiency in the other direction from TPC to ITS is constant over the occupancy shows that the problem is with the TPC, not with ITS or the general conditions.
It can only mean that a some TPC tracks are either not reconstructed or are of poor quality such that they cannot be matched to ITS.
The situation is further investigated in Fig.~\ref{fig:tpcits} (b), which shows the lowest TPC pad row of the reconstructed TPC tracks matched to ITS.
While normally it is assumed that most matchable TPC tracks should have clusters in the inner pad rows (e.\,g.~pad row $< 20$), a significant fraction of TPC tracks lacks such clusters.
This was identified as a problem during the seeding, where seeds are found, but discarded during the ambiguity solving step since in the high occupancy region shared clusters are stolen by better, longer tracks, such that some track segments in the inner pad rows are discarded.

For the time being, this is mitigated by loosening several cuts:
the minimum pad row allowed for matching a TPC track is raised to 80.
The allowed fraction of shared clusters of TPC track segments is increased from 10\% to 20\%.
And the track following during the merging stage, which was foreseen to pick up missing short track segments in adjacent sectors, is repurposed to propagate all tracks also from the middle of a sector without checking for shared clusters.
This improves the ITS-TPC matching, and increases e.\,g.~the~$\text{K}_{\text{S}}^0$ yield significantly ($\sim$2x in high IR Pb--Pb), but comes at the cost of significantly increased fake tracks and fake cluster attachment rates ($\sim$50\% higher) as well as shared clusters ($\sim$2x higher).

Another problem is that the three-iteration refit of long, looping, low~$p_{\text{T}}$ tracks often fails.
The track merging connects all mergeable legs of such tracks, and the track fit tries to follow the full helix.
Originally, this was considered advantageous since it enables the usage of the full information, in particular more clusters are available for~$\text{d}E/\text{d}x$ calculation.
However, meanwhile the additional legs are discarded during the compression~\cite{bib:ctd2019}.
And the track fit often fails when the track parameters are mirrored and propagated from one leg to the next, particularly for very low~$p_{\text{T}}$ tracks with large uncertainty due to the small number of clusters per leg.
In addition, following the full looping code adds more branches to the code and increases warp divergence during GPU processing reducing the GPU reconstruction speed.

\section{Tracking Improvements}
\label{section2}

Several improvements to the TPC tracking algorithms have been and are being investigated.
The implementation to the GPU code started in 2025 and is split into three phases.

Phase~I disables merging different legs of looping tracks, but splits the full helix in one track per leg.
These individual tracks are processed independently, keeping the relation such that secondary legs can be removed later, or their clusters could be used for~$\text{d}E/\text{d}x$.
In addition, the cluster assignment criterion during the ambiguity solving of the sector tracking has been optimized.
In case of multiple tracks competing for a cluster, more priority is given to the track with smaller global~$\chi^2$, reducing the weight of the track length.
Shared clusters are counted starting at the outermost pad row, which allows for a larger number of shared clusters in the inner pad rows for long tracks.
Finally, the interpolation-based cluster rejection is improved.
For this, it is necessary to move the interpolation to one fit iteration earlier.
This is possible now, since independently a sector track refit at the beginning of the track merging was added, guaranteeing reasonable initial parameters.
Now, position and covariance are stored in the first, inward iteration.
The interpolation and the~$\chi^2$ cut are performed in the second, outward iteration.
Cluster rejection happens later in the third iteration.
This approach solves the instability issue mentioned before, where the track fit cannot recover if the parameters drift away from the real trajectory.
Note that while this is more stable, it is still not ideal, but an iterative approach to redo the fit after every rejected cluster is computationally infeasible.

Phase~II aims to improve the cluster attachment, by allowing additional attachment during the fit.
Furthermore, the track is practically rebuilt during iterations 2 and 3 of the fit, i.\,e.~the cluster-to-track attachment is performed again from scratch.
Instead of performing a~$\chi^2$ cut for cluster rejection in the second iteration, the fast cluster search grid from the sector tracking~\cite{bib:tns} is queried for all clusters compatible with the~$\chi^2$ cut and all candidates together with their residuals are written to temporary storage.
When iteration 2 has finished for all tracks, a second ambiguity solving step is executed.
The significantly lower number of tracks compared to initial seeds allows for a much more sophisticated algorithm with multiple candidates and iterations.
As in the sector tracking, candidates are assigned to the track with the best combination of global~$\chi^2$ and track length and optionally other parameters allowing a configurable amount of cluster sharing.
However, tracks failing to get their best candidate attached are still eligible for the second candidate, and so on.
Only after this final cluster-to-track attachment, the third fit iteration is executed on the new tracks.
To avoid running more iterations of the fit, it performs a simple~$\chi^2$ cut for final cluster rejection without interpolation.

Phase~III extends the track rebuilding of phase~II by extrapolating all tracks inwards and outwards, also over sector boundaries, aiming to find missing track segments particularly in the inner pad rows.
The cluster candidates found during the extrapolation are handled by the same ambiguity solving, optionally using stricter cuts compared to clusters found by interpolation.
In addition, the seeding of the sector tracking will be improved to run multiple iterations on the leftover clusters after ambiguity solving.
Phase~III is expected to show the largest improvements for cluster attachment in the inner pad rows, so in some sense it is the most important phase.
But technically, it depends on phase~I and~II, so it is implemented last.

\begin{figure}[!t]
  \centering
  \includegraphics[width=0.94\linewidth]{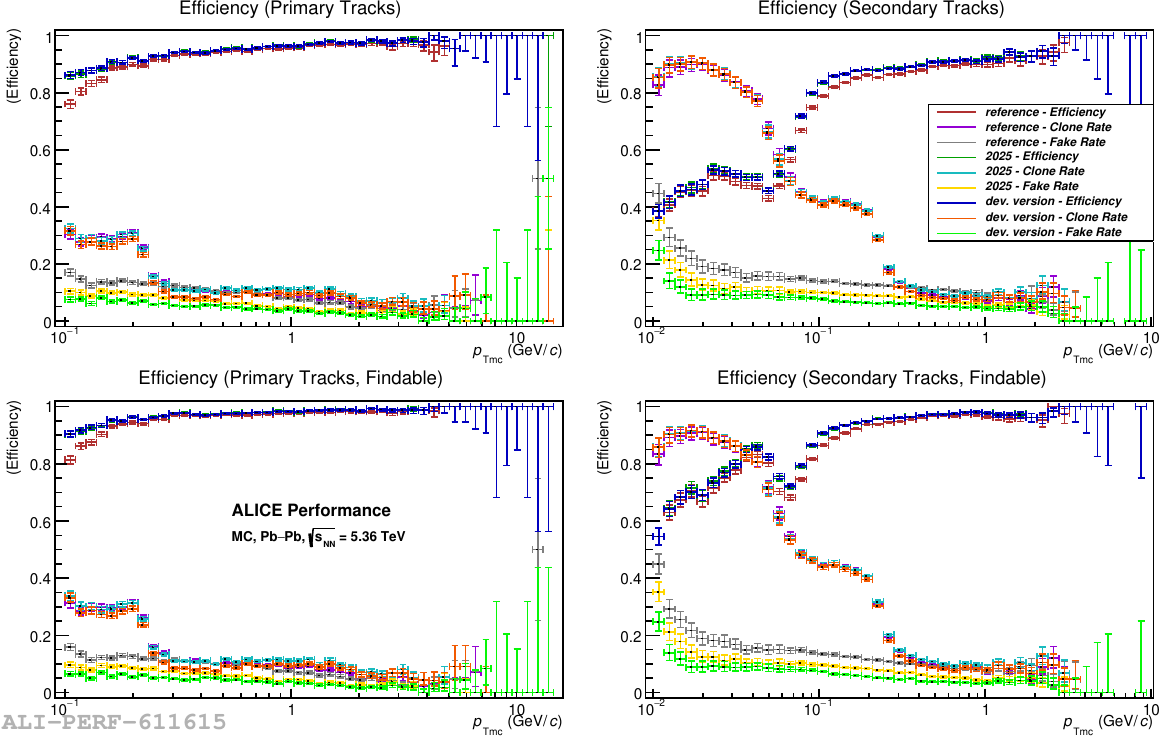}
  \caption{Track finding efficiency, clone rate and fake rate comparison of the original ALICE TPC tracking algorithm in operation until August 2025, the new version including the phase~I improvements deployed in September 2025, and the development version containing the phase~II improvements.
  Findable tracks are required to have at least 70 hits in the TPC, while the efficiency for general tracks is computed for tracks that have at least as many hits in the TPC as the~$p_{\text{T}}$-dependent~$n_{\text{Cl}}$ cut during the tracking requires.}
  \label{fig:eff1}
\end{figure}

Phase~I improvements have been fully implemented and merged in the ALICE tracking code in September 2025.
Phase~II improvements are fully implemented but not yet optimized for memory consumption and speed.
For phase~III, only a first development version exists implementing the extrapolation feature.
Fig.~\ref{fig:eff1} shows that phase~I yields significantly higher efficiency at much lower fake rates compared to the original version, particularly at low~$p_{\text{T}}$, while phase~II maintains the efficiency while reducing the fake rate further.

Current development focuses primarily onthe  extrapolation feature.
At the time of writing, it increases the number of clusters attached to tracks significantly by 5.4\% (increasing the total fraction of attached clusters from 60.1\% to 63.3\%).
However, the additional 5.4\% of clusters exhibit a much higher relative fake contribution of 5.7\% than the clusters prior to extrapolation.
The total fake attachment rate increases from 1.28\% to 1.51\%.
It is assumed that this can be further improved through special handling of cases where the extrapolation is overly aggressive.
A set of heuristic cuts is currently under development to abort the extrapolation in cases with a high probability of fake attachment.
For comparison, the current loose cut mitigation results in an even higher fake attachment rate of 1.78\%.

\section{Cluster Reduction}
\label{section3}

TPC online compression is based on online cluster finding, entropy reduction and ANS (Asymmetric numeral systems) encoding, and finally the removal of clusters not used for physics (\cite{bib:ctd2019} gives an overview).
In a nutshell, the cluster removal first protects all clusters that are in a protective tube around physics tracks.
Then, it removes all unprotected clusters in a removal tube around secondary legs of looping tracks and around very low-$p_{\text{T}}$ tracks, which are not used for physics.
It is crucial that the online TPC cluster removal does not remove any clusters of physics tracks, since data discarded online can never be recovered.
This can only occur when a pyhsics track reconstructed offline is not found online.
In that case, its clusters are not protected, so clusters can be removed if they are in the vicinity of a looping track by coincidence.

\begin{figure}[!htb]
  \centering
  \subfloat[]{\includegraphics[width=0.475\linewidth]{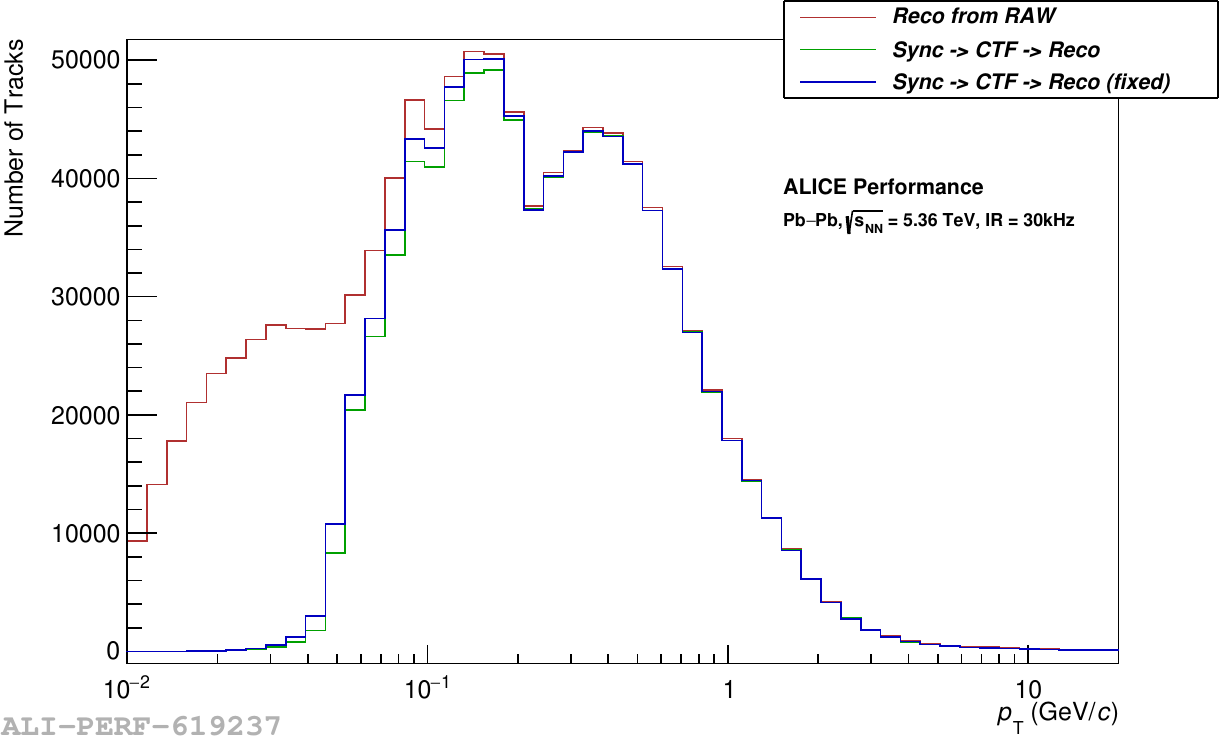}}
  \qquad
  \subfloat[]{\includegraphics[width=0.475\linewidth]{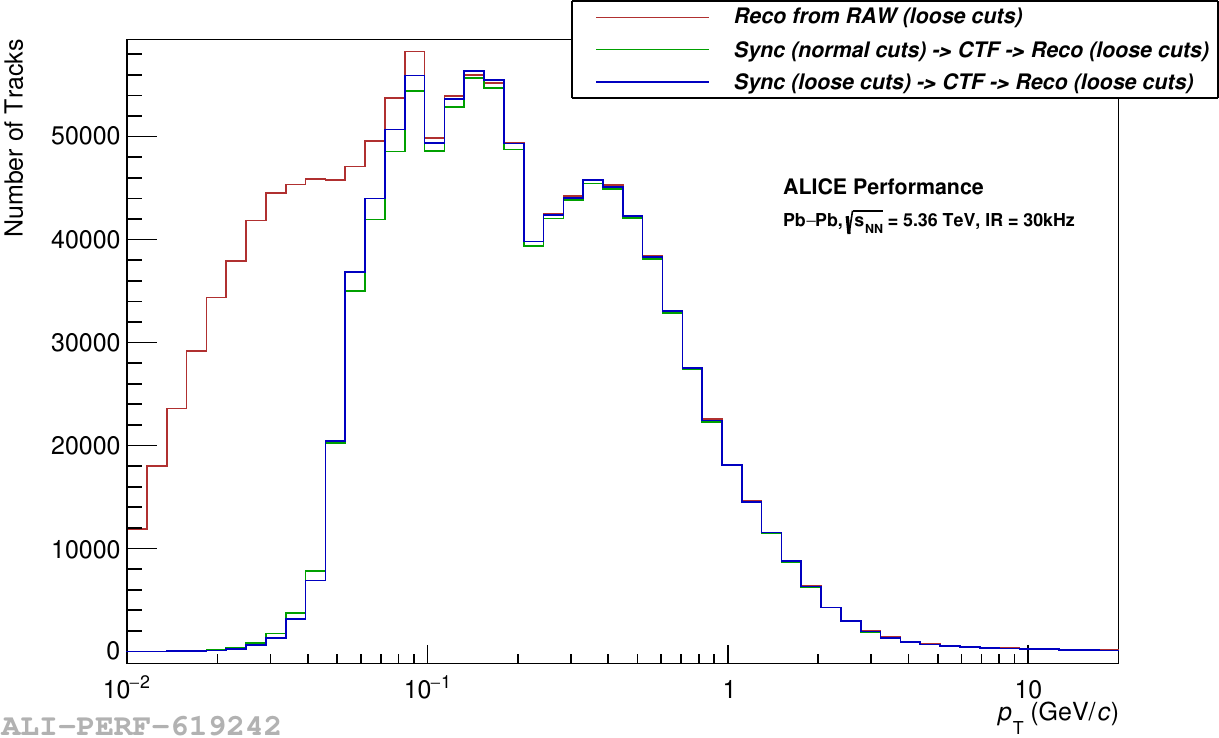}}
  \caption{Track~$p_{\text{T}}$ spectrum comparison of direct reconstruction from raw data to first compression to a CTF and then reconstruction from CTF.
  The tracking algorithm including the phase~I improvements is used.
  (a) applies strict cuts during both the online and offline phases, and compares the reconstruction via CTF in the original form and with the fix to protect the history of the tracks.
  (b) applies loose cuts during the reconstruction, and compares the usage of strict cuts versus loose cuts during the compression to CTF.}
  \label{fig:ncls}
\end{figure}

Therefore, significant changes to the tracking algorithm mandate an analysis of the cluster rejection.
The impact of the compression to CTF is compared by reconstructing some time frames directly from raw data and comparing that to the case where the same time frames are first compressed to CTF with online cluster removal, and then reconstructed from CTF.
Fig.~\ref{fig:ncls} (a) shows a small loss of tracks above 200 MeV/c.
The loss of tracks below 200 MeV/c is expected and corresponds to the looping track removal.
The loss above 200 MeV/c appears since the ``history'' of the track was not protected.
Here, history means clusters that were involved in the seeding of that track, but are not attached to the final track (e.\,g.~since they could have been attached to another track instead).
This effect must have been present but suppressed before, but due to the better tracking of, and cluster attachment to, low~$p_{\text{T}}$ looping tracks in the phase~I improvements, more looping tracks are removed and the loss became more pronounced.
While this was easy to fix by protecting the track history, Fig.~\ref{fig:ncls} (b) shows a more severe issue.
If the offline reconstruction uses looser cuts to mitigate the cluster attachment at inner pad rows, the same must be done during the online reconstruction.
Otherwise parts of tracks that are not found online can be removed if the cluster is attached to a nearby looping track.

\begin{figure}[!htb]
  \centering
  \includegraphics[width=0.87\linewidth]{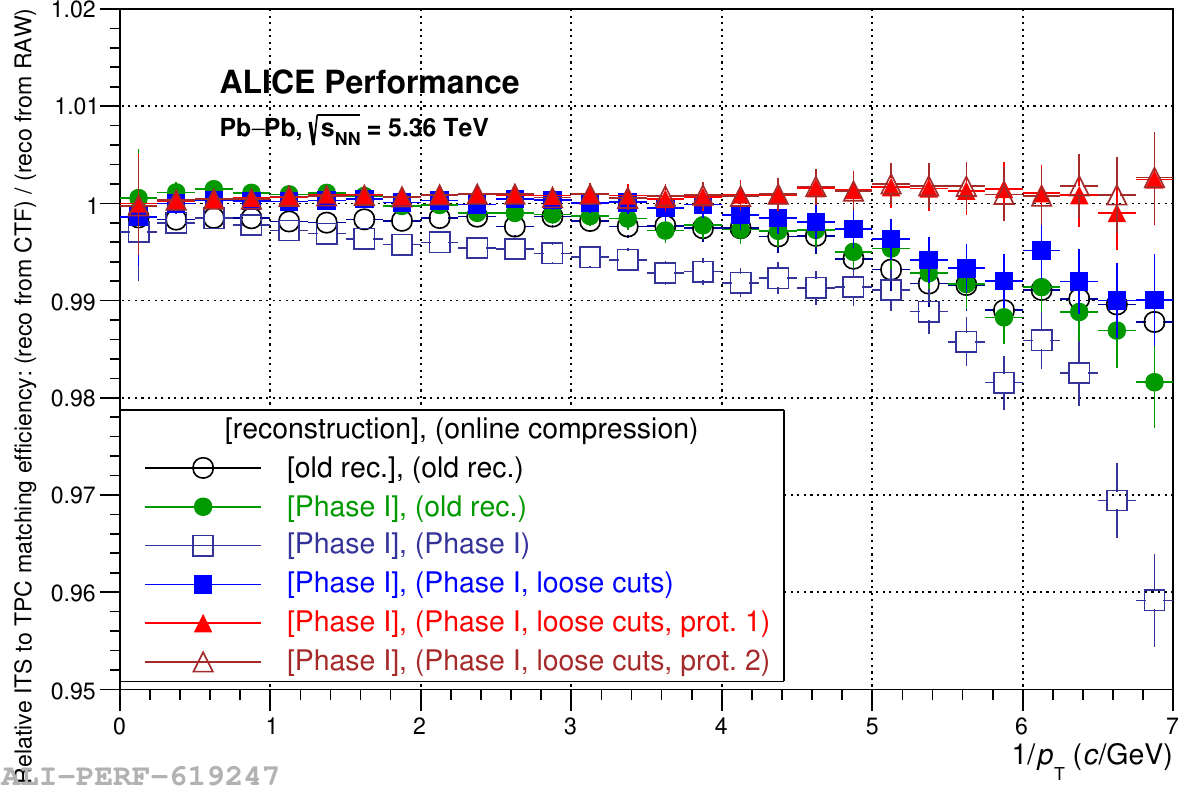}
  \caption{Fraction of TPC tracks found given an ITS track in reconstruction from CTF relative to direct reconstruction from the raw data.
  Square brackets in the legend indicate whether the original reconstruction or phase~I improvements were used, in any case with loose cuts.
  Brackets indicate which cuts and which reconstruction version was used for the online phase to produce the CTF.}
  \label{fig:ruben}
\end{figure}

In order to increase the safety margin for the future, in particular in view of the better finding of low~$p_{\text{T}}$ tracks with the phase~I improvements, a more detailed study was conducted and additional protections have been implemented.
Protection 1 increases the protective tube size around physics tracks preventing cluster removal, while reducing the tube size around looping legs and low~$p_{\text{T}}$ tracks where unprotected clusters are removed.
Protection 2 performs a dynamic protection on top, further enlarging and reducing the tube sizes for clusters in the~20 innermost pad rows, at the sector edges, and in regions of high local occupancy.
Fig.~\ref{fig:ruben} shows the effect of the cluster removal on the ITS-TPC matching with different cuts, tracking versions, and additional protection.
The effect is most pronounced for low~$p_{\text{T}}$ tracks below 250 MeV/c.
Comparing the dark blue (phase~I) to the black curve (old tracking) shows that it has become much more important to apply the correct, loose, safe cuts online, otherwise a significant fraction of tracks is lost.
While applying loose cuts online is principally enough (light blue curve), the extra protections (light red and dark red curves) yield additional improvements down to very low~$p_{\text{T}}$.
In particular, the curves with the extra protections go consistently above 1, i.\,e.~the matching becomes slightly more efficient after the online compression than in reconstruction from the raw data.
The reason is that the tracking works slightly better in the cleaned up environment after cluster removal.

\section{GPU Framework Improvements and Usage of GPUs for Online and Offline Processing}

Since the start of online data taking with GPUs in the beginning of Run 3, the ALICE GPU framework and reconstruction have been improved significantly, in particular also in view of using it for offline reconstruction.
In order to run on all GPUs with optimal parameters (like number of threads and blocks or shared memory sizes), an automatic tuning framework based on Optuna\footnote{https://optuna.org/ - A hyperparameter optimization framework}  has been established.
This has improved the online processing performance by~10\% in 2025, compared to the tuned values which were last updated in 2020.
Two very important framework features to note are RTC (Run Time Compilation) and a deterministic mode~\cite{bib:chep2024}.
RTC enables the recompilation of the GPU code at runtime, applying a couple of optimizations, in particular replacing runtime constants by constexpr variables removing branches, and in case of online reconstruction completely removing code paths only used for offline, also reducing branching.
Today this yields a performance improvement of 20\% to 30\%.
In addition, it enables on-the-fly compilation for architectures not enabled in the global build available from CVMFS.
The deterministic mode is a combination of runtime and compile time options that guarantees a fully deterministic processing and comparable results from CPU and GPU at the bit level including all intermediate processing steps.
This has proven extremely valuable for debugging and validation and has even revealed several GPU compiler bugs.

\looseness=-1 Since 2023, ALICE offloads 50\% to 60\% of the offline reconstruction to GPUs achieving a 2x to 2.5x speedup~\cite{bib:chep2023}.
Porting more steps to GPUs will increase the GPU utilization and speed up offline reconstruction on the EPNs.
Today, also the TPC track model decoding (see~\cite{bib:ctd2019}) runs on GPUs for a 1.2\% to 3\% speedup~\cite{bib:gabrielelhcp}.
More importantly, ITS tracking on GPUs has been validated at the end of 2025, and shall be used for the reconstruction of 2025 Pb--Pb data in 2026 with a speedup of 26\% compared to running only TPC on the GPU.

\section{Conclusions}

ALICE employs GPUs heavily to speed up online and offline processing.
Today, 99\% of online reconstruction and~$\sim$60\% of offline processing runs on GPUs.
The offline speedup is up to 2.5x, and should increase by 26\% for the 2025 Pb--Pb reconstruction with also ITS tracking on GPU.
Finally, a speedup of 5x is aimed for, offloading the full barrel tracking which constitutes 80\% of the offline processing workload.

Some TPC tracking inefficiencies have been identified, mostly stemming from the fact that the tracking originates from the HLT implementation which gives more priority to speed than to efficiency.
This is amplified by the increased occupancy in the TPC compared to the estimations from the technical design report.
Three phases of improvements have been foreseen, with the first being already deployed, the second being tested right now, and the third being in development.
A significant improvement in tracking efficiency, reduction in fake rate and fake cluster attachment, and an increase in the number of clusters per track has already been demonstrated.

The ALICE GPU framework is in constant development.
Notable new features with a broad range of applications are RTC and the deterministic mode.

\end{document}